\documentclass[sigconf,screen,authorversion]{acmart}

\AtBeginDocument{%
  }

\copyrightyear{2026}
\acmYear{2026}
\setcopyright{cc}
\setcctype{by}
\acmConference[FSE Companion '26]{34th ACM Joint European Software Engineering Conference and Symposium on the Foundations of Software Engineering}{July 05--09, 2026}{Montreal, QC, Canada}
\acmBooktitle{34th ACM Joint European Software Engineering Conference and Symposium on the Foundations of Software Engineering (FSE Companion '26), July 05--09, 2026, Montreal, QC, Canada}
\acmDOI{10.1145/3803437.3805548}
\acmISBN{979-8-4007-2636-1/2026/07}

\usepackage{cleveref}
\usepackage[most]{tcolorbox}

\begin{document}

\title{Reproducible, Explainable, and Effective Evaluations of Agentic AI for Software Engineering}

\author{Jingyue Li}
\email{jingyue.li@ntnu.no}
\orcid{0000-0002-7958-391X}
\affiliation{%
  \institution{Norwegian University of Science and Technology}
  \city{Trondheim}
  \country{Norway}
}

\author{Andr\'e Storhaug}
\email{andre.storhaug@ntnu.no}
\orcid{0000-0002-5321-7196}
\affiliation{%
  \institution{Norwegian University of Science and Technology}
  \city{Trondheim}
  \country{Norway}
}

\begin{abstract}
With the advancement of Agentic AI, researchers are increasingly leveraging autonomous agents to address challenges in software engineering (SE). However, the large language models (LLMs) that underpin these agents often function as black boxes, making it difficult to justify the superiority of Agentic AI approaches over baselines. Furthermore, missing information in the evaluation design description frequently renders the reproduction of results infeasible. To synthesize current evaluation practices for Agentic AI in SE, this study analyzes 18 papers on the topic, published or accepted by ICSE 2026, ICSE 2025, FSE 2025, ASE 2025, and ISSTA 2025. The analysis identifies prevailing approaches and their limitations in evaluating Agentic AI for SE, both in current research and potential future studies. To address these shortcomings, this position paper proposes a set of guidelines and recommendations designed to empower reproducible, explainable, and effective evaluations of Agentic AI in software engineering. In particular, we recommend that Agentic AI researchers make their Thought–Action–Result (TAR) trajectories and LLM interaction data, or summarized versions of these artifacts, publicly accessible. Doing so will enable subsequent studies to more effectively analyze the strengths and weaknesses of different Agentic AI approaches. To demonstrate the feasibility of such comparisons, we present a proof‑of‑concept case study that illustrates how TAR trajectories can support systematic analysis across approaches.  
  
\end{abstract}
    



\maketitle

\section{Introduction}
Agentic AI technologies are increasingly being adopted by software engineering (SE) researchers. At ICSE 2025, only seven research track papers included the keyword ``agent'' in their titles. In contrast, the number of accepted papers for ICSE 2026 featuring ``agent'' in their titles has risen to 30\footnote{\url{https://conf.researchr.org/track/icse-2026/icse-2026-research-track} (accessed 12 January 2026)}. Although Agentic AI technologies have demonstrated superior performance in addressing SE challenges compared to classical approaches, such as static and dynamic code analysis, and machine learning techniques, including deep neural networks, the underlying reasons for this superiority are often difficult to justify and reproduce. This challenge stems from the black-box nature of large language models (LLMs) and the inherent randomness in their outputs. Furthermore, querying LLMs for comprehensive evaluations can be computationally expensive. As Agentic AI approaches gain popularity, they are likely to become standard baselines for future studies. Consequently, it is imperative to establish methods for evaluating Agentic AI approaches in a reproducible, explainable, and efficient manner.

To address these challenges, we first analyzed a set of papers \cite{ICSE2025-1, ICSE2025-2, ICSE2025-3, ICSE2025-4, ICSE2025-5, ICSE2025-6, ICSE2025-7, ICSE2026-1, ICSE2026-2, ICSE2026-3, ICSE2026-4, ISSTA2025-1, ISSTA2025-2, ISSTA2025-3, ASE2025-1, ASE2025-2, ASE2025-3, ASE2025-4} published or accepted in recent prestigious software engineering conferences, including ICSE 2025, ICSE 2026, FSE 2025, ISSTA 2025, and ASE 2025. The selection of papers is based on their relevance and availability at the time of this paper’s submission. For conferences with published proceedings by the time of submitting this paper, such as ICSE 2025, FSE 2025, and ISSTA 2025, we selected papers containing the keyword ``agent'' in their titles. For ICSE 2026 and ASE 2025, whose proceedings have not yet been published, we selected papers from the accepted paper lists and included only those with preprints available as of 12 January 2026. This paper selection is not intended to be exhaustive or fully representative. Rather, the goal is to obtain a preliminary understanding of the current state of empirical evaluations in SE studies employing Agentic AI approaches, identify potential challenges, and propose guidelines for future improvements. Results of the analysis show that:
\begin{itemize}
    \item Most current evaluations of Agentic AI approaches compare these methods against classical techniques, deep learning models, or naive LLMs as baselines. Only one out of the 18 investigated papers included comparisons with relevant state-of-the-art Agentic AI baselines.
    \item Many studies employing Agentic AI approaches conducted ablation experiments to assess the contribution of individual components to overall performance.
    \item Several studies performed failure analyses and case studies to gain deeper insights into the factors influencing the success or failure of their Agentic AI approaches.  
    \item   A subset of studies incorporated cost analyses to provide evidence regarding the cost-effectiveness of their proposed Agentic AI approaches. 
\end{itemize}

The analysis also indicates that several parameters can significantly influence evaluation outcomes, including the versions of the LLMs, temperature configurations, prompt templates, and the inherent randomness of LLM outputs. Although some studies have conducted experiments to examine the impact of these parameters, the majority have not. Accordingly, this study proposes guidelines, strategies, and methodologies to enhance future evaluation of Agentic AI in software engineering:

\begin{itemize}
    \item publishing the prompts used, temperature configurations, and LLM versions for \textbf{improving reproducibility}. 
    \item making the TAR trajectories and LLM interaction data, or summarized versions of these artifacts, openly available and automatically analyzed to enable \textbf{explainable, cost‑effective comparisons across different Agentic AI approaches}.
\end{itemize}

To illustrate the practicality of using TAR data for comparing approaches, we presented an example method that analyzes agents’ TAR trajectories per case and then performs cross‑case comparisons as meta‑analysis. In contrast to the classical approach, where studies evaluate agents on specific benchmarks independently and publish only the aggregate results, our proposed method enables more fine‑grained, explainable,  and cost‑effective comparisons. 

The remainder of this paper is organized as follows. Section \ref{data-analysis} describes the process used to identify and analyze the chosen papers and results. Section \ref{proposal} presents our proposed approach and provides examples. Section \ref{discussion} discusses our ideas. Finally, Section \ref{conclusion} concludes the paper and outlines future work directions.

\section{Data Collection and Analysis}
\label{data-analysis}

As explained, the purpose of our analysis of existing papers applying Agentic AI for software engineering is to gain a preliminary understanding of the current state of practice in evaluating proposed approaches. Based on the title containing the keyword ``agent'' and the availability of papers in the proceedings or preprints by January 12, 2026, we analyzed the following papers from recent top software engineering conferences. 

\begin{itemize}
    \item ICSE 2025 \cite{ICSE2025-1, ICSE2025-2, ICSE2025-3, ICSE2025-4, ICSE2025-6, ICSE2025-7}. 
    \item ICSE 2026 (the accepted papers with preprints available) \cite{ICSE2026-1, ICSE2026-2, ICSE2026-3}.
    \item FSE 2025 \cite{FSE2025-1}.
    \item ISSTA 2025 \cite{ISSTA2025-1, ISSTA2025-2, ISSTA2025-3}. 
    \item ASE 2025 (the accepted papers with preprints available) \cite{ASE2025-1, ASE2025-2, ASE2025-3}.  
\end{itemize}

For each paper, the study identified and summarized its research focus, the Agentic AI approaches employed, the evaluation design, the baselines used, and the analyses conducted to enhance the reproducibility, explainability, and effectiveness of the evaluations. The results are categorized and synthesized as follows. 

\textbf{Research focuses.} These papers aim at addressing different software engineering challenges, including: testing or auditing \cite{ICSE2025-1, ICSE2025-3, ICSE2025-4, ICSE2025-5, ICSE2026-1, ICSE2026-2, ISSTA2025-2, ISSTA2025-3}, software design \cite{ICSE2025-2}, program repair \cite{ICSE2025-6, ISSTA2025-1}, code generation \cite{ICSE2025-7}, error analysis \cite{ICSE2026-3}, bug reproducing \cite{ICSE2026-4, ASE2025-2}, automated building \cite{FSE2025-1}, automated specification generation \cite{ASE2025-1}, and effort estimation \cite{ASE2025-3}. The broad scope of software engineering topics covered highlights the relevance and growing interest in applying Agentic AI approaches, underscoring the need for comprehensive studies to evaluate these methods. 

\textbf{Agentic AI approaches applied.} Eleven of the eighteen studies investigated \cite{ICSE2025-1, ICSE2025-2, ICSE2025-3, ICSE2025-4, ICSE2026-1, ICSE2026-2, ICSE2026-3, ISSTA2025-2, ASE2025-1, ASE2025-2, ASE2025-3} employed multi-agent approaches, while the remaining studies focused on developing and applying a single agent to address software engineering problems. 

\textbf{Baselines.} Six of the investigated studies \cite{ICSE2025-1, ICSE2025-5, ICSE2025-6, ICSE2026-1, ISSTA2025-1, ASE2025-2} employed classical approaches as baselines and assessed performance differences using well-established benchmark datasets. One study \cite{ASE2025-3} compared an Agentic AI approach with deep learning methods. To evaluate Agentic AI approaches, several studies compared their performance against naïve prompt-based LLMs \cite{ICSE2025-2, ICSE2025-7, ICSE2026-4, ISSTA2025-2, FSE2025-1} or fine-tuned models \cite{ICSE2025-3}. In \cite{ICSE2026-2}, which focuses on agent development, the evaluation centered on comparing the performance of multiple LLMs underlying the agents. For studies addressing novel challenges without prior work \cite{ICSE2025-4, ISSTA2025-3, ASE2025-1}, baselines included human-curated datasets, system execution logs, or expert judgments. Among all the investigated studies, only one \cite{ICSE2026-3} adopted an existing Agentic AI approach as a baseline, indicating that comparative evaluations of such methods are still in their infancy and require further research to establish guidelines and best practices. 

\textbf{Extra studies performed in the evaluation.} Beyond performance comparisons between Agentic AI approaches and their baselines, the most common evaluation technique was ablation analysis. Thirteen of the eighteen studies conducted ablation analyses to examine the contributions of individual agents or tools within the Agentic AI pipeline.     
 To enhance the explainability of the results, several studies \cite{ICSE2026-3, ICSE2026-4, ISSTA2025-2, ASE2025-1, ASE2025-2} have \textit{analyzed the failures} of the proposed approaches to better understand the agents’ challenges and limitations. Other works \cite{ICSE2025-1, ICSE2025-6, ISSTA2025-2, ASE2025-2} present \textit{case studies} that illustrate why Agentic AI approaches succeed in practice.
To improve reproducibility, most studies provide \textit{prompt examples or raw data}, although not all do so. Because the temperature setting of LLMs can influence their outputs, \citet{ICSE2025-2} and \citet{ICSE2025-7} conducted \textit{LLM temperature sensitivity analyses}, demonstrating that different temperature configurations yield varying results. Furthermore, \citet{ICSE2025-1, ICSE2025-5}, and \citet{ ICSE2026-4} reported the \textit{specific temperature settings} used in their experiments. 
To mitigate the randomness in LLM outputs during evaluations, studies such as \cite{ICSE2025-1, ASE2025-1} perform \textit{multiple repetitions} and report averaged results. 
While all investigated studies disclose the names of the LLMs employed, only a few \cite{ICSE2025-7, ISSTA2025-1, ASE2025-1} provide \textit{precise version identifiers}, such as GPT-3.5-0125 in \cite{ISSTA2025-1}.
To demonstrate the efficiency of Agentic AI approaches, \textit{analyses of time complexity and API costs} were conducted in \cite{ICSE2025-2}, while \textit{LLM token costs} were examined in \cite{ICSE2026-1, ISSTA2025-1, ISSTA2025-3}.
 
\section{Evaluation Guideline and Strategy Proposals}
\label{proposal}

The data analysis results underscore several weaknesses in current practices for evaluating Agentic AI approaches. To enhance the reproducibility, explainability, and efficiency of evaluations in the era of Agentic AI for software engineering, this study proposes the following guidelines, strategies, and methodologies for discussion.

\textbf{Reproducibility} Several existing studies have attempted to mitigate the limitations of LLMs’ black-box nature and output randomness by publishing the \textbf{prompts used, temperature configurations, and LLM versions}. However, not all papers adhere to these practices. Collecting and publishing such information introduces minimal overhead and can significantly enhance the reproducibility of the evaluations. \textbf{Thus, reporting the values of these parameters should be established as a standard requirement within the software engineering community for Agentic AI approaches.} 

\textbf{Explainability}. While ablation studies, failure analyses, and examples of success offer some insight into the strengths and weaknesses of proposed Agentic AI approaches, they are insufficient to provide a systematic and comprehensive explanation of their superiority over baseline methods. Because Agentic AI approaches are perceived as novel, researchers often emphasize their positive aspects. However, \citet{FSE2025-2} presented evidence suggesting that agent-based approaches may not outperform agentless alternatives. Given the early stage of Agentic AI research, there are currently few Agentic AI approaches that serve as state-of-the-art baselines. As the popularity of Agentic AI continues to grow, it is essential to develop innovative methodologies that systematically assess the strengths and weaknesses of these approaches in comparative evaluations. 

Inspired by \cite{ASE2025-4}, which compared three Agentic AI approaches using agents’ Thought-Action-Result (TAR) trajectories and LLM interactions, we propose to \textbf{enforce open access to agent TAR trajectories} and \textbf{develop methods for automatically analyzing these trajectories} to effectively improve the explainability of Agentic AI evaluations. Since TAR trajectories can be automatically generated as execution logs, creating them introduces minimal overhead. Publishing these trajectories as open-access data alongside research papers would enable future researchers to better understand why a proposed approach achieves its observed performance. Moreover, when multiple Agentic AI approaches aim to address the same problem, automated analysis and comparison of their trajectories could provide highly interpretable insights into the relative strengths and weaknesses of the approaches under evaluation. Currently, trajectory analysis remains a manual process \cite{ASE2025-4}. Thus, developing automated techniques for such analyses represents a valuable direction for future research in evaluating Agentic AI approaches for software engineering. The study by \cite{Li-2026} demonstrated that LLMs can accurately summarize the intent of malicious code. As a result, leveraging summarization models \footnote{\url{https://huggingface.co/docs/transformers/en/tasks/summarization}} to automatically analyze TAR trajectories may represent a promising solution. 

\textbf{Efficiency} Evaluating Agentic AI approaches can be costly, particularly when interaction with commercial LLM APIs is required. Although running evaluations multiple times to report averaged results is considered good practice, such repetition imposes high additional costs for researchers attempting to reproduce the findings. Instead of relying solely on replicating baselines and comparing performance metrics (such as accuracy, false positives, and false negatives, etc.), which may encourage cherry-picking and remain susceptible to LLM output randomness, a more robust and efficient alternative is for \textbf{both baseline and subsequent approaches to publish their TAR trajectory data}. This approach would enable qualitative comparison and justification of the strengths and weaknesses of the methods under evaluation. If such comparisons can be automated, the cost of assessing baselines against newly proposed approaches could be significantly reduced. This, in turn, would encourage researchers to focus on developing novel and valid ideas rather than spending excessive time reproducing the results of baseline approaches, which, in practice, may not always be fully reproducible.

To demonstrate the idea, we provide an example of such an automated comparison. We use the openly available TAR trajectories \footnote{\url{https://huggingface.co/datasets/andstor/favia_trajectories}} from \cite{storhaug2026faviaforensicagentvulnerabilityfix} that uses an agent to identify CVE (Common Vulnerabilities and Exposures) vulnerability fixes. They run their experiments using different models, including Qwen3-235B, Llama-3.3-70B-Instruct, and Gemma-3-27B. Each model represents an agent. To illustrate how TAR trajectories can be leveraged to analyze differences in agent performance, we compare a subset of agents’ TAR trajectories across the three models.
We select TAR trajectories from 10 randomly sampled failed runs of vulnerability‑fix detection using the Qwen3‑235B model and compare them with the corresponding TAR trajectories from agents using the Gemma‑3‑27B and Llama‑3.3‑70B‑Instruct models on the same 10 runs. Although the agent using the Qwen3‑235B model failed to detect the correct vulnerability fixes in these 10 runs, the agents powered by the Gemma‑3‑27B and Llama‑3.3‑70B‑Instruct models succeeded in detecting some of them. This highlights the importance of understanding the strengths and limitations of different agents when they are applied to the same task. 

\begin{figure*}[htbp]
    \centering
    \includegraphics[width=\textwidth]{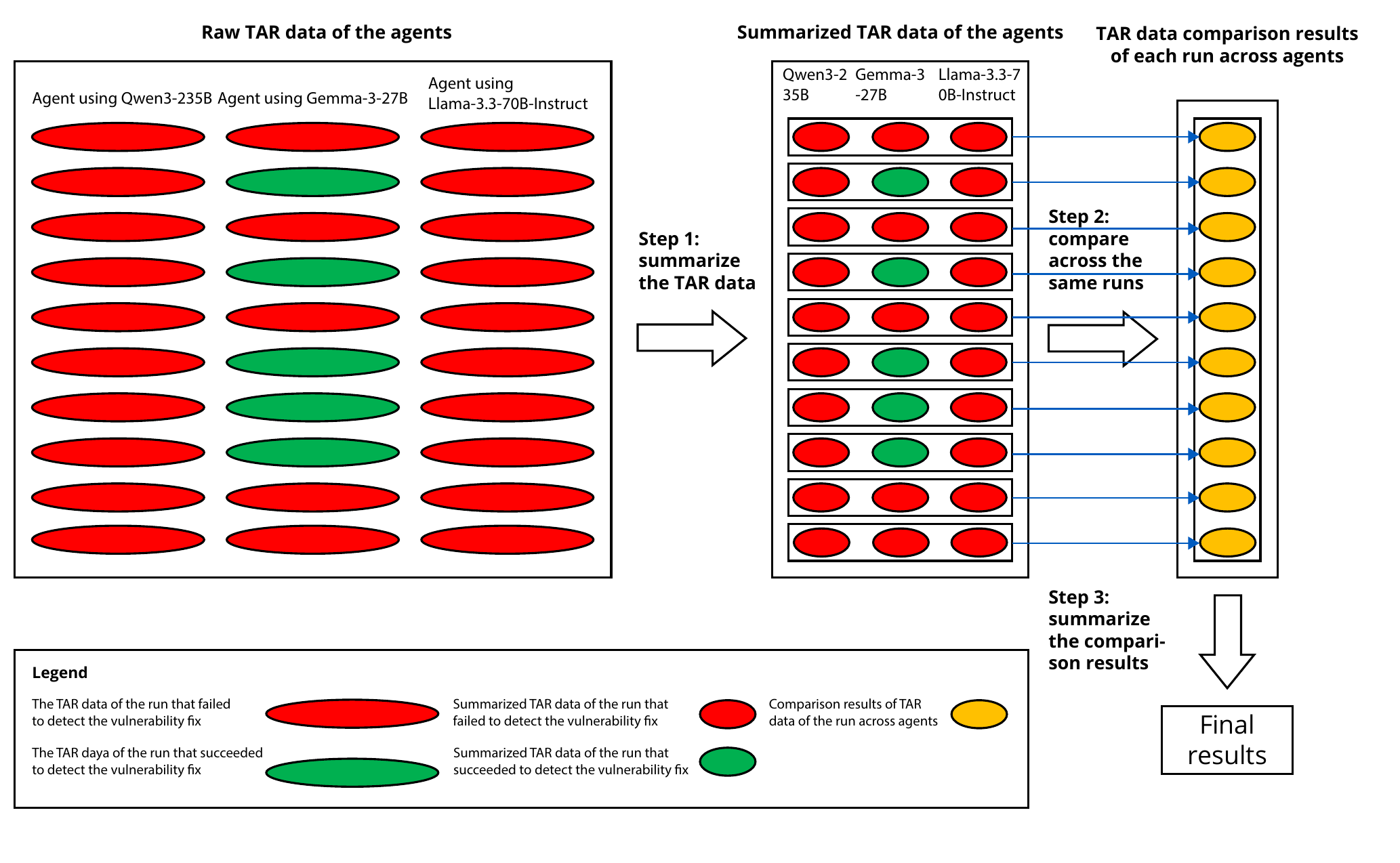}
    \caption{The process to analyze TAR trajectories}
    \label{fig:analysis-approach}
\end{figure*}

To automatically analyze and compare the TAR trajectories, we use Kimi K2.5 Instant \cite{kimiteam2026kimik25visualagentic}. One challenge in analyzing TAR trajectories is that they can be very long. To address this challenge, we applied multi-step summarization to manage their size and complexity. As shown in \Cref{fig:analysis-approach}, starting from raw TAR trajectories, we automatically performed three-step analyses: 
\begin{itemize}
    \item Step 1: summarize individual runs
    \item Step 2: generate comparative analyses between the agents across the same runs
    \item Step 3:  aggregate these comparisons to extract recurring strengths and weaknesses
\end{itemize}

\begin{figure}[ht]
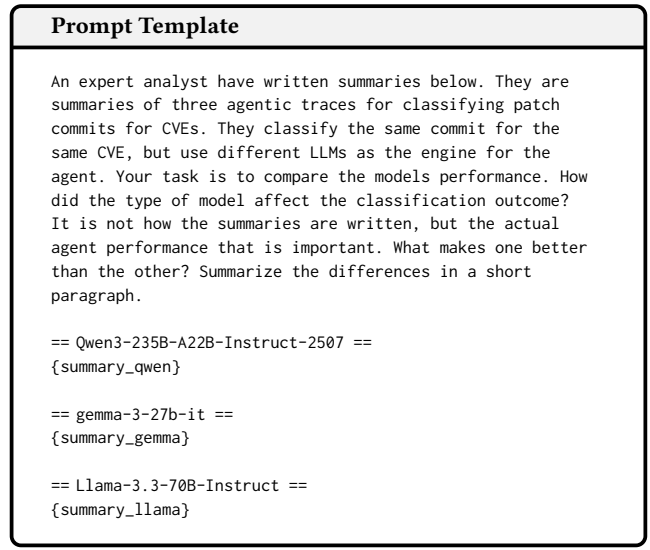

\begin{tcblisting}{
    title=Prompt Template,
    enhanced,
    colbacktitle=black!5!white,
    coltitle=black,
    colback=white,
    colframe=black!75!black,
    boxrule=0.5mm,
    fonttitle=\bfseries,
    fontupper=\ttfamily,
    arc=1mm,
    breakable=false,
    listing only,
    listing options={
        aboveskip=0pt,
        belowskip=0pt,
        numbers=none,
        basicstyle=\ttfamily\footnotesize,
        frame=none,
        xleftmargin=0pt,
        xrightmargin=0pt,
        columns=flexible,
        breaklines=true,
        breakatwhitespace=true,
        breakautoindent=false,
        breakindent=0pt,
        lineskip=0pt,
        escapechar=§,
        literate={\{}{\{}{1}%
                {\}}{\}}{1}%
                {=}{{=}}{1}%
      }
}
An expert analyst have written summaries below. They are summaries of three agentic traces for classifying patch commits for CVEs. They classify the same commit for the same CVE, but use different LLMs as the engine for the agent. Your task is to compare the models performance. How did the type of model affect the classification outcome? It is not how the summaries are written, but the actual agent performance that is important. What makes one better than the other? Summarize the differences in a short paragraph.

== Qwen3-235B-A22B-Instruct-2507 ==
{summary_qwen}

== gemma-3-27b-it ==
{summary_gemma}

== Llama-3.3-70B-Instruct ==
{summary_llama}
\end{tcblisting}
\caption{Prompt template for comparative analyses between the agents on the same runs.}
\label{fig:prompt-template-2}
\end{figure}

\begin{figure}[ht]
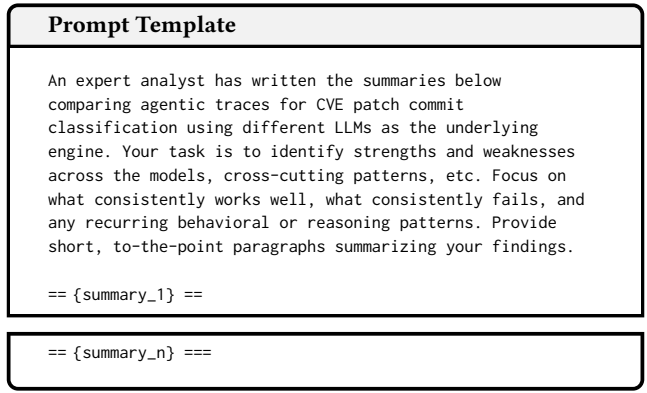

\begin{tcblisting}{
    title=Prompt Template,
    enhanced,
    colbacktitle=black!5!white,
    coltitle=black,
    colback=white,
    colframe=black!75!black,
    boxrule=0.5mm,
    bottom=0pt,
    sharp corners=south,
    fonttitle=\bfseries,
    fontupper=\ttfamily,
    arc=1mm,
    breakable=false,
    listing only,
    listing options={
        aboveskip=0pt,
        belowskip=0pt,
        numbers=none,
        basicstyle=\ttfamily\footnotesize,
        frame=none,
        xleftmargin=0pt,
        xrightmargin=0pt,
        columns=flexible,
        breaklines=true,
        breakatwhitespace=true,
        breakautoindent=false,
        breakindent=0pt,
        lineskip=0pt,
        escapechar=§,
        literate={\{}{\{}{1}%
                {\}}{\}}{1}%
                {=}{{=}}{1}%
      }
}
An expert analyst has written the summaries below comparing agentic traces for CVE patch commit classification using different LLMs as the underlying engine. Your task is to identify strengths and weaknesses across the models, cross-cutting patterns, etc. Focus on what consistently works well, what consistently fails, and any recurring behavioral or reasoning patterns. Provide short, to-the-point paragraphs summarizing your findings.

== {summary_1} ==
\end{tcblisting}
\begin{tcblisting}{
    enhanced,
    colbacktitle=black!5!white,
    coltitle=black,
    top=0pt,
    colback=white,
    colframe=black!75!black,
    boxrule=0.5mm,
    fonttitle=\bfseries,
    fontupper=\ttfamily,
    arc=1mm,
    breakable=false,
    sharp corners=north,
    listing only,
    listing options={
        aboveskip=0pt,
        belowskip=0pt,
        numbers=none,
        basicstyle=\ttfamily\footnotesize,
        frame=none,
        xleftmargin=0pt,
        xrightmargin=0pt,
        columns=flexible,
        breaklines=true,
        breakatwhitespace=true,
        breakautoindent=false,
        breakindent=0pt,
        lineskip=0pt,
        escapechar=§,
        literate={\{}{\{}{1}%
                {\}}{\}}{1}%
                {=}{{=}}{1}%
      }
}
== {summary_n} ===
\end{tcblisting}
\caption{Prompt template for aggregate analysis of comparisons 1 to n.}
\label{fig:prompt-template-3}
\end{figure}

 The prompts we used to analyze the TAR trajectories are shown in \Cref{fig:prompt-template-1} (for Step 1), \Cref{fig:prompt-template-2} (for Step 2), and \Cref{fig:prompt-template-3} (for Step 3). The results of the analysis are shown in \Cref{fig:analysis-example}. 

\begin{figure}[ht]
\begin{tcblisting}{
    title=Prompt Template,
    enhanced,
    colbacktitle=black!5!white,
    coltitle=black,
    colback=white,
    colframe=black!75!black,
    boxrule=0.5mm,
    bottom=0pt,
    fonttitle=\bfseries,
    fontupper=\ttfamily,
    sharp corners=south,
    arc=1mm,
    breakable=false,
    listing only,
    listing options={
        aboveskip=0pt,
        belowskip=0pt,
        numbers=none,
        basicstyle=\ttfamily\footnotesize,
        frame=none,
        xleftmargin=0pt,
        xrightmargin=0pt,
        columns=flexible,
        breaklines=true,
        breakatwhitespace=false,
        breakautoindent=false,
        breakindent=0pt,
        lineskip=0pt,
        escapechar=§,
        literate={\{}{\{}{1}%
                {\}}{\}}{1}%
                {=}{{=}}{1}%
      }
}
Analyze the following trace of a security analysis session and provide a small summary of why the agent { 'succeeded' if resolved else 'failed'} the task.

== {role} at iteration 0 ==
{trace_0}
\end{tcblisting}

\begin{tcblisting}{
    enhanced,
    colbacktitle=black!5!white,
    coltitle=black,
    colback=white,
    colframe=black!75!black,
    boxrule=0.5mm,
    top=0pt,
    fonttitle=\bfseries,
    fontupper=\ttfamily,
    arc=1mm,
    breakable=false,
    sharp corners=north,
    listing only,
    listing options={
        aboveskip=0pt,
        belowskip=0pt,
        numbers=none,
        basicstyle=\ttfamily\footnotesize,
        frame=none,
        xleftmargin=0pt,
        xrightmargin=0pt,
        columns=flexible,
        breaklines=true,
        breakatwhitespace=true,
        breakautoindent=false,
        breakindent=0pt,
        lineskip=0pt,
        escapechar=§,
        literate={\{}{\{}{1}%
                {\}}{\}}{1}%
                {=}{{=}}{1}%
      }
}
== {role} at iteration {n} ==
{trace_n}
\end{tcblisting}
\caption{Prompt template for summarizing individual run from iteration 0 to n.}
\label{fig:prompt-template-1}
\end{figure}

\begin{figure*}[ht]
\begin{tcblisting}{
    title=Aggregated Analysis,
    left=2pt, right=2pt,
    enhanced,
    colbacktitle=black!5!white,
    coltitle=black,
    colback=white,
    colframe=black!75!black,
    boxrule=0.5mm,
    fonttitle=\bfseries,
    fontupper=\ttfamily,
    arc=1mm,
    breakable=false,
    listing only,
    listing options={
        aboveskip=0pt,
        belowskip=0pt,
        numbers=none,
        basicstyle=\ttfamily\footnotesize,
        frame=none,
        xleftmargin=0pt,
        xrightmargin=0pt,
        columns=flexible,
        breaklines=true,
        breakatwhitespace=true,
        breakautoindent=false,
        breakindent=0pt,
        lineskip=0pt,
        escapechar=§,
        literate={\{}{\{}{1}%
                 {\}}{\}}{1}%
                 {→}{-> }{1}%
      }
}
## Cross-Model Analysis: Agentic CVE Classification

### What Consistently Works Well

**Temporal and contextual verification** emerges as the strongest success pattern. Llama-3.3-70B consistently outperforms others by checking foundational facts first-dates, versions, CVE metadata-before diving into code analysis. This "verify then analyze" discipline prevents the rabbit holes that trap other models. When Llama succeeds, it's because it treats timeline impossibilities and version mismatches as hard stops rather than details to work around.

**Structured tool use with recovery** also works reliably. Gemma-3-27b demonstrates that rigorous, step-by-step tool application (CVE retrieval → code search → evidence synthesis) produces accurate results when executed properly. Llama's ability to recover from initial tool errors and still complete its analysis shows that resilience matters as much as capability.

### What Consistently Fails

**Pattern matching without validation** is the dominant failure mode. Both Qwen3 and Gemma repeatedly fixate on superficial keyword overlaps-"CSRF" in message and CVE, "SQL injection" near "escaping," "normalize" implying security-without verifying functional relevance. This creates false confidence: models assign 5/5 ratings while missing six-year temporal gaps or conflating entirely different vulnerability classes (TLS padding vs. UTF-8 encoding).

**Tool abandonment under pressure** appears across traces. When tools fail or return errors, models don't pause-they hallucinate. Qwen3 fabricates CVE details when retrieval fails; others skip required verification steps entirely. The availability of tools doesn't guarantee their use when reasoning shortcuts seem available.

**Infinite examination loops** plague Qwen3 specifically. Despite strong analytical frameworks, it gets trapped iterating without conclusion, suggesting that reasoning depth without task adherence is worse than shallow but directed analysis.

### Recurring Behavioral Patterns

**Epistemic humility varies dramatically**. Llama regularly acknowledges uncertainty and cross-checks evidence; Qwen and Gemma exhibit overconfidence bias, pressing forward with conclusions despite contradictory signals. This isn't about model size-Qwen's 235B parameters don't prevent basic logical errors like examining the wrong file version.

**Architectural reasoning gaps** appear consistently. Models struggle to trace data flow from vulnerability entry points to modified code, often analyzing commits in isolation rather than verifying they address the actual CVE attack surface. Gemma shows flashes of coverage awareness (checking all seven vulnerable paths), but this rarely translates to execution.

**Misdirection by commit messages** is a systemic weakness. Models repeatedly trust commit summaries over code reality, leading to backwards logic (interpreting relaxed controls as added security) or dismissal of relevant changes (ignoring path normalization as "just Windows handling").

### Key Differentiator

The decisive factor is **analytical discipline over reasoning sophistication**. Llama's 70B parameters outperform Qwen's 235B not through deeper reasoning chains, but through stricter adherence to verification protocols: check dates, check versions, check affected files, check tool outputs. Sophisticated analysis of the wrong question-Qwen's hallmark-produces confident wrong answers. Methodical checking of the right questions produces correct ones, even with imperfect technical explanations.
\end{tcblisting}
\caption{Final results of the TAR trajectory analysis}
\label{fig:analysis-example}
\end{figure*}

The results show that automated trajectory analysis can surface structured, system-specific insights, including: 1) Agents' strengths, e.g., \textit{``Llama-3.3-70B consistently outperforms others by checking foundational facts, first dates, versions, CVE metadata, before diving into code analysis.''} 2) The weaknesses of the agents, e.g., \textit{``Pattern matching without validation is the dominant failure mode.''} 3) Their key differentiator, e.g., \textit{``Llama's 70B parameters outperform Qwen's 235B not through deeper reasoning chains, but through stricter adherence to verification protocols: check dates, check versions, check affected files, check tool outputs.''} The results illustrate the feasibility of using LLM-based summarization to analyze TAR trajectories to produce interpretable, qualitative evaluations of agent behavior at scale. 

\section{Discussions}
\label{discussion}
When comparing Agentic AI approaches for software engineering tasks, researchers often evaluate different agents using the same benchmark datasets in baseline studies. Our example suggests that comparing TAR trajectories directly could serve as an alternative strategy that reduces the cost and effort of rerunning baseline approaches. However, this requires baseline studies to make their TAR trajectories openly available—an expensive requirement given that TAR trajectories can be extremely large.
A more practical alternative is for baseline studies to release only the summaries of their TAR trajectories for each run. In this setup, researchers execute  Step 1 of \Cref{data-analysis} locally to generate TAR trajectories summaries, which are then shared as open-access artifacts. Comparative evaluations of Agentic AI approaches can then focus on Steps 2 and 3 in \Cref{data-analysis}, enabling efficient, scalable analysis without requiring full TAR trajectories. One potential threat is the possibility of deceptive or biased summarized TAR outputs. Given the studies' prompts, temperature settings, LLM version, and summarization prompts, any biases introduced during summarization can be examined. 

If a study aims to compare multiple Agentic AI approaches using benchmark datasets that differ from those used in the baseline papers, relying solely on TAR trajectories or their summaries may yield misleading conclusions, as agent behavior can vary significantly across datasets. In such cases, baseline studies should provide sufficient detail to support reproducibility of agent behavior, including prompts, temperature configurations, and the specific LLM versions employed. Nevertheless, making the TAR trajectories, or their summarized forms, open access can still help researchers effectively compare agents’ behavior across different datasets and better understand their dataset-dependent characteristics. One potential threat to validity arises when using agents to analyze agent trajectories, as this may introduce bias into the results. To mitigate this risk, employing multiple agents to cross‑validate the findings can help improve the overall reliability of the analysis. 

\section{Conclusion and Future Work}
\label{conclusion}
As software engineering researchers increasingly adopt Agentic AI approaches to address complex challenges, ensuring that evaluations of these approaches are reproducible, explainable, and efficient is essential. This paper first analyzes recent high-quality studies investigating Agentic AI in software engineering and summarizes their evaluation designs and practices. The analysis identifies several issues related to current evaluation methods. Based on these findings, this study proposes guidelines and strategies, particularly leveraging TAR trajectories, to enhance the reproducibility, explainability, and efficiency of evaluations. To demonstrate the feasibility of LLM-based summarization for automatically analyzing TAR trajectories, we present a case study. As future work, we plan to conduct more extensive empirical studies to compare Agentic AI approaches in software engineering, aiming to develop more systematic and effective methods for analyzing TAR trajectories. 

\section{Data Availability}
\label{data-availability}

The data from our experiments are available at \url{https://github.com/andstor/agentic-ai-eval-replication-package}.

\bibliographystyle{ACM-Reference-Format}
\bibliography{references}

\end{document}